\documentclass[a4paper]{article}

\usepackage{INTERSPEECH2022}
\usepackage[colorinlistoftodos,prependcaption]{todonotes}
\usepackage{makecell}
\usepackage{url}
\usepackage{amsmath}
\usepackage{color}
\usepackage[hidelinks]{hyperref}
\usepackage{booktabs}
\usepackage{svg}
\usepackage{pifont}
\newcommand{\cmark}{\ding{51}}%
\newcommand{\xmark}{\ding{55}}%
\usepackage{multirow}
\usepackage{enumitem}

\makeatletter
\newcommand{\printfnsymbol}[1]{%
  \textsuperscript{\@fnsymbol{#1}}%
}
\makeatother

\usepackage{math_defs}

\title{Defense against Adversarial Attacks on Hybrid Speech Recognition using \\ Joint Adversarial Fine-tuning with Denoiser}
\name{Sonal Joshi,
        Saurabh Kataria\printfnsymbol{1} \thanks{\printfnsymbol{1}equal contribution.\\This work is supported by DARPA projects: GARD (www.darpa.mil/ program/guaranteeing-ai-robustness-against-deception) and RED (www.darpa.mil/program/reverse-engineering-of-deceptions)},
        Yiwen Shao\printfnsymbol{1}, \\
        Piotr {\.Z}elasko,
        Jes{\'u}s  Villalba,
        Sanjeev Khudanpur,
        Najim Dehak}

\address{Center for Language and Speech Processing, Johns Hopkins University, Baltimore, MD}
\email{\{sjoshi12,skatari1,yshao18,pzelask2,jvillal7,khudanpur,ndehak3\}@jhu.edu}

\begin{document}

\maketitle
\begin{abstract}
Adversarial attacks are a threat to automatic speech recognition (ASR) systems, and it becomes imperative to propose defenses to protect them. In this paper, we perform experiments to show that K2 conformer hybrid ASR is strongly affected by white-box adversarial attacks. We propose three defenses--denoiser pre-processor, adversarially fine-tuning ASR model, and adversarially fine-tuning joint model of ASR and denoiser. Our evaluation shows denoiser pre-processor (trained on offline adversarial examples) fails to defend against adaptive white-box attacks. However, adversarially fine-tuning the denoiser using a tandem model of denoiser and ASR  offers more robustness. We evaluate two variants of this defense--one updating parameters of both models and the second keeping ASR frozen. The joint model offers a mean absolute decrease of 19.3\% ground truth (GT) WER with reference to baseline against fast gradient sign method (FGSM) attacks with different $L_\infty$ norms. The joint model with frozen ASR parameters gives the best defense against projected gradient descent (PGD) with 7 iterations, yielding a mean absolute increase of 22.3\% GT WER with reference to baseline; and  against PGD with 500 iterations, yielding a mean absolute decrease of 45.08\% GT WER and an increase of 68.05\% adversarial target WER.
\end{abstract}


\section{Introduction}

In today's world, voice-based smart assistants are ubiquitous--be it using phones, dedicated home assistants like Alexa, Google Home, Apple Pod, or as call center agents~\cite{talebi2021megamind}. A
core technology behind these assistants is automatic speech recognition (ASR) whose goal is to transcribe speech to text. Recent work~\cite{carlini-18,Qin2019,wang2020adversarial} has shown that ASR systems are vulnerable to adversarial inputs, which contain specially crafted mostly human inaudible noise. Considering the threat of these adversarial attacks, it becomes of foremost importance to propose defensive countermeasures to protect ASRs. These countermeasures broadly fall into three categories--pre-processing, stochastic, and adversarial training. Pre-processing defenses intend to remove the adversarial noise from the signal before before passing it into the machine learning system~\cite{yuan-18,yang2018characterizing,esmaeilpour2021}. Stochastic defenses introduce randomness into the model. Thus the model used to craft the adversarial sample differs slightly (but stochastically) from the model used to evaluate the sample, reducing attack effectiveness. Randomized smoothing is the most common stochastic defense~\cite{cohen2019certified}. Finally, adversarial training tries to make the model inherently robust by training on dynamically generated adversarial examples~\cite{madry2018towards,jati_adversarial_2020}.
The major contributions of this work are highlighted below:
\begin{itemize}[leftmargin=*]
    \item We evaluate the robustness of a strong baseline K2 Conformer hybrid ASR model  against  white-box attacks, i.e. when the adversary is aware of parameters of the model. To the best of our knowledge, this is the first work that fully utilizes a differentiable hybrid ASR model to study adversarial robustness. 
    \item We propose four defenses--a pre-processing time-domain denoiser, adversarial fine-tuning and two variants of joint adversarial training of a pre-processing denoiser with ASR model.
    \item We evaluate these defenses against strong adaptive white-box attacks i.e. when the adversary is aware of parameters of defense model along with those of ASR.
\end{itemize}
The rest of the paper is as follows. In Section \ref{sec:advASR}, we introduce adversarial attacks on ASR. 
In Section \ref{sec:defenses} , we describe defenses; followed by experiments and results in Section \ref{sec:exp} and \ref{sec:results} respectively. 
\begin{table*}
\caption{Table describing types of transcripts with shorthand symbol used and example}
\resizebox{1\textwidth}{!}{
\begin{tabular}{|l|c|c|c|c|}
\toprule
\textbf{Description }& \textbf{Shorthand} & \textbf{Under attack?} & \textbf{Example} \\
\midrule
What is the actual (human transcribed) content of speech signal? & actual & \xmark & This is the human ASR output \\
What is the text that the ASR system predicts ? & benign & \xmark & Thus is the real ASR output\\
What is the text that the adversary wants to achieve? & target & \cmark & Transfer \$1000 from my account\\
What is the text that is actually predicted by ASR after the attack? &  adversarial & \cmark & Transfer is sand from my account\\
\bottomrule
\end{tabular}
\label{tab:transcripts}
}
\end{table*}

\begin{table*}
\caption{Table describing different Word Error Rate (WER) metrics used for evaluation of successful defense}
\label{tab:metrics}
\resizebox{1\textwidth}{!}{
\begin{tabular}{|l|c|l|c|}
\toprule
\textbf{Description} & \textbf{WER type} & \makecell{\textbf{Formula} \\ WER($<ref>,<hyp>$)} & \makecell{\textbf{Defense} \\ \textbf{success}} \\
\midrule
Does the defense harm the un-attacked system? & Benign ground truth & Benign = WER(actual,benign) & $\downarrow$ \\
Did the attacker succeed in denial of service? & Adversarial ground truth & GT = WER(actual,adversarial)  & $\downarrow$\\
Did the adversary make the system recognize what he/she wants? & Adversarial target & TGT=WER(target,adversarial) & $\uparrow$ \\
\bottomrule
\end{tabular}
}
\end{table*}
\section{Adversarial attacks on ASR}
\label{sec:advASR}

An ASR system can be considered as a function $\hat{\yvec}=f(\xvec,\theta)$, which predicts a sequence of words $\hat{\yvec}$ given a audio waveform $\xvec$. $f$ is an statistical model described by a set of parameters $\theta$. ASR systems are known to be vulnerable to adversarial attacks~\cite{carlini-18}. The attacker adds a small perturbation to the benign signal to alter the prediction of the system. Depending on the goals of the attacker, we find different attack modalities.
When an adversarial example fools the ASR into predicting a particular target phrase that an adversary desires, it is called a \emph{targeted} adversarial attack. {\em Untargeted} attacks, by comparison, simply induce transcription errors and are not of as much concern in the ASR context~\cite{Qin2019}. Table~\ref{tab:transcripts} explains the concept of targeted attack via an example. Suppose we have a speech signal $\mathbf{x}$. Without any attack, a human will transcribe $\mathbf{x}$ as \texttt{\textbf{actual} = This is the human ASR output}. Now using $\mathbf{x}$ as input to the ASR model, the output transcription by the ASR is denoted as \texttt{\textbf{benign} = Thus is the real ASR output}. Suppose, the adversary wants the ASR model to predict the target phrase denoted by \texttt{\textbf{target} = Transfer \$1000 from my account}. To achieve this goal, he/she crafts an adversarial example $\mathbf{x'}$  such that when $\mathbf{x'}$ is given as input to the ASR, the model produces the output transcription \texttt{\textbf{adversarial} = Transfer a sand from my account}. One can find Word Error Rate (WER) between the pairs of the transcriptions as shown in Table~\ref{tab:metrics}. Let WER between reference ($ref$) and hypothesis ($hyp$) be denoted by $WER(ref,hyp)$. $Benign=WER(actual,benign)$ denotes the Benign ground truth WER, which measures the performance of the ASR system in non-attack conditions. A good ASR system, and hence defense, should have Benign WERs as low as possible, indicated by $\downarrow$ in the column \textit{Defense success}. $GT=WER(actual,adversarial)$ is called the adversarial ground truth WER. GT is a performance indicator of how much attacker succeeded in denial-of-service i.e. introducing untargeted spelling errors. An higher value indicates as successful untargeted attack while a low value is characteristic of a robust ASR. $TGT=WER(target,adversarial)$ denotes the adversarial target WER and is performance indicator whether the adversary was successful in getting the ASR predict the chosen target phrase. While the adversary wants TGT WER to as low as possible (meaning, target phrase to be recognised perfectly), an ideal defense will make it be as high as possible. This is indicated by $\uparrow$ in the column \textit{Defense success}.\\
\textbf{Attack Algorithms:} An adversarial example is computed as $\mathbf{x'} = \mathbf{x} + \mathbf{\delta}$ where  is
$\mathbf{x}$ is a benign signal and $\mathbf{\delta}$ is a small adversarial perturbation. Many attack algorithms in the literature~\cite{wang2020adversarial} propose different ways to compute $\mathbf{\delta}$. 
In this work, we consider FGSM~\cite{Goodfellow2015} and Projected Gradient Descept (PGD) attacks~\cite{madry2018towards}.
For targeted attacks, PGD optimizes delta by gradient descent iterations that minimizes minimizing the ASR loss $L$ between the target phrase selected by the attacker $\mathbf{y}^\mathrm{target}$ and the adversarial transcript predicted by ASR. Thus for iteration $i+1$,
\begin{align}
\mathbf{\delta_{i+1}} = \mathrm{clip}_\varepsilon( \mathbf{\delta}_{i} - \alpha \mathrm{sign}(\nabla L(f(\mathbf{x},\theta),\mathbf{y}^\mathrm{target}))\;.
\end{align}
Throughout this paper, PGD-i indicates the number of iterations used for PGD attack (eg: PGD-7 means 7 iterations). At every iteration, $\clip$ function (projection) ensures that $\lVert \mathbf{\delta}\rVert_{\infty} \le \varepsilon$, keeping the attack imperceptible. 
We choose the learning rate $\alpha$ at every iteration is one fifth of the max-norm. FGSM is single iteration version of PGD with step $\alpha=\varepsilon$. While attacking any system (with or without defense), we assume that it fully white-box and adaptive, meaning the adversary knows not only the speech recognition model parameters but also the defense. This is the worst-case scenario to evaluate robustness, a commonly expected norm for evaluation of defenses as it exposes the systems weakest links~\cite{tramer2020adaptive}.
\begin{table*}
\centering
\caption{\label{tab:wer_results}
Ground Truth (GT) Word error rate (\%) ($\downarrow$) for K2 ASR systems under FGSM and PGD-7 attacks and defenses. \textit{RS$\sigma$} stands for randomized smoothing with a $\sigma$ parameter.}
\resizebox{\textwidth}{!}{
\begin{tabular}{@{}lccccccccccc@{}}
\toprule
\textbf{System}  & \textbf{Benign}  & \multicolumn{5}{c}{\textbf{FGSM Attack}} & \multicolumn{5}{c}{\textbf{PGD-7 Attack}} \\
\cmidrule(r){1-1}\cmidrule(lr){2-2}\cmidrule(lr){3-7}\cmidrule(lr){8-12}
$L_{\infty}$ (max-norm) &  & 0.0001 & 0.001 & 0.01 & 0.1 & 0.2  & 0.0001 & 0.001 & 0.01 & 0.1 & 0.2  \\
\midrule
Baseline (full LibriSpeech test-clean) & 4.34 & 5.12 &	7.13 &	10.62 &	70.27 &	90.80 &	6.29 &	22.38 &	63.50 &	95.24 &	97.68 \\
\midrule
Baseline (reduced LibriSpeech test-clean) & \bf 4.53	 & 5.36	& 7.22 & 	11.56 & 	74.84 & 	91.32 & 	6.44  & 	25.89  & 	63.63  & 	95.71  & 	97.90 \\
\quad + \textit{RS0.001} & 4.66 & \bf 4.88 & 10.43 & 13.12 & 79.23 & 94.49 & \bf 4.83 & 10.48 & 65.68 & 94.73 & 97.17 \\
\quad + \textit{RS0.001} + \textit{DENOISER} & 4.95 & 4.97 & 5.95 &	21.65 &	81.72 &	98.78 &	4.97 &	5.90 &	40.61 &	92.64 &	95.76 \\
\textit{ADV-FINETUNE-ASR} + \textit{RS0.001} & 4.73 & \bf 4.88 & 
\bf 5.70 & 20.48 & 82.30 & 99.27 & 4.88 & 5.90 & 40.42 & 93.71 & 96.10 \\
\textit{ADV-FINETUNE-JOINT} + \textit{RS0.001} & 5.22 & 5.07 & \bf 5.70 & \bf 6.00 & \bf 21.40 & \bf 55.58 & 5.27 & \bf 5.22 & 12.19 & 78.99 & 92.74\\
\textit{ADV-FINETUNE-JOINT-ASRfrozen} + \textit{RS0.001} & 5.64 & 6.09 & 6.24 & 9.65 & 90.05 & 100.00 & 6.24 & 6.39 & \bf 10.29 & \bf 65.29 & \bf 89.81 \\
\bottomrule
\end{tabular}
}
\end{table*}

\section{Defenses}
\label{sec:defenses}
\subsection{Randomized smoothing} 
Randomized smoothing is a stochastic defense that adds random, normally-distributed noise with standard deviation $\sigma$  to the input. This additive normal noise tries to mask the gradients that are essential in computing adversarial examples. If $\sigma$ is too high, the benign accuracy  reduces and hence it is vital to find $\sigma$ that offers robustness without reducing the accuracy. Previous work on speaker identification defenses~\cite{joshi2021adversarial} show that this defense can be easily combined with other defenses and may offer additional protection against high $L_\infty$ attacks.
\subsection{Adversarial fine-tuning of ASR model} This defense is a variant of adversarial training~\cite{madry2018towards}. Instead of full adversarial training, which leads to convergence issues in ASR, we propose to bootstrap from a pre-trained ASR (which is trained using clean/benign examples as normally done) and then fine-tune using the model using PGD adversarial attacks. We call this model \textit{ADV-FINETUNE-ASR}. For an ASR model denoted by $f(\cdot,\theta)$, where $\theta$ are the model parameters, adversarial training is done by minimizing the loss function given by 
\begin{align}
\label{eq:class_advtrn1}
    \theta^{*} = \argmin_\theta \Expcond{ J(f(\mathbf{x}+\delta^{*},\theta),\mathbf{y})}{(\mathbf{x},\mathbf{y})\sim \mathcal{D}} \;,
\end{align}
where $J$ is the ASR loss function (lattice-free MMI in our case), $\mathcal{D}$ is the set of training audio-transcript pairs $(\mathbf{x},\mathbf{y})$, and $\delta^{*}={\argmax}_{\delta, \left\|\delta\right\|_\infty\le \varepsilon}J(f(\mathbf{x}+\delta,\theta),\mathbf{y})$ is adversarial perturbation optimized by PGD iterations.
\subsection{Denoiser} 
The pre-processing denoiser defense maps adversarial signals to benign.
The denosier was trained using deep regression approach \cite{xu2014regression} in time-domain.
Training objective function $\mathcal{L}_{\text{sup}}$ is Multi-Resolution Short-Time Fourier Transform (MRSTFT) auxiliary loss. 

\begin{align}
\label{eq:mrstft}
\mathcal{L}_{\text{sup}} &= \mathbb{E}_{\mathbf{x,x'}\sim P_{\mathcal{B},\mathcal{A}}}[\sum_{m=1}^M \mathcal{L}_{\text{sup}}^{(m)}(\mathbf{x},\mathbf{x'})]\;,\\
    \mathcal{L}_{\text{sup}}^{(m)}(\mathbf{x},\mathbf{x'}) &=  \mathcal{L}_{\text{sc}}^{(m)}(\mathbf{x},g(\mathbf{x'}, \phivec)) 
    + \mathcal{L}_{\text{mag}}^{(m)}(\mathbf{x},g(\mathbf{x'}, \phivec))),\\
    \mathcal{L}_{\text{sc}}^{(m)}(\mathbf{x},\mathbf{\hat{x}}) &= \frac{\||\operatorname{STFT}^{(m)}(\mathbf{x})|-|\operatorname{STFT}^{(m)}(\mathbf{\hat{x}})|\|_{F}}{\||\operatorname{STFT}^{(m)}(\mathbf{x})|\|_{F}},\\
    \mathcal{L}_{\text{mag}}^{(m)}(\mathbf{x},\mathbf{\hat{x}}) &= \frac{1}{N} \| \log |\operatorname{STFT}^{(m)}(\mathbf{x})| \notag\\&- \log |\operatorname{STFT}^{(m)}(\mathbf{\hat{x}})| \|_1\;;
\end{align}
where $\mathbf{x}$ is a benign signal and $\mathbf{x'}$ is the corresponding adversarial signal,  $\mathbf{\hat{x}}=g(\mathbf{x'},\phivec)$ is the predicted benign and $\phivec$ are the parameters of the denoiser. 
$\mathcal{B}$ and $\mathcal{A}$ denote the benign and adversarial domains and $P_{\mathcal{B},\mathcal{A}}$ denotes their joint distribution, which is obtained from a dataset of offline attack samples. $\mathcal{L}_{\text{sup}}$ uses $M$ different STFT with different frame-shift and frame-lengths, which are indexed by $m=1\dots M$.
The number of time-frequency bins in the STFT are denoted by $N$, while $||\cdot||_F$ refers to the Frobenius matrix norm. 
\subsection{Adversarial fine-tuning of joint ASR and Denoiser model} The disadvantage of using denoiser as pre-processor is that, in a fully white-box scenario, the adversary can break the system by backpropogating through the combined denoiser+ASR network and computing adaptive adversarial attacks--i.e., attacks that adapt to the defense (albeit at the expense of higher computing cost). Therefore, to make the denoiser itself robust to adaptive white-box attacks, 
we propose to adversarially fine-tune the pre-trained denoiser in tandem with the ASR model using on-the-fly PGD attacks, trying to minimize the ASR cross-entropy. We call this model \textit{ADV-FINETUNE-JOINT}. Similar to standard adversarial training in~\eqref{eq:class_advtrn1}, the optimum parameters for denoiser and ASR models are given by
\begin{align}
\label{eq:class_advtrn2}
    \theta^{*}, \phi^{*} = \argmin_{\theta, \phi}\Expcond{ J(f(g(\mathbf{x}+\delta^{*},\phi),\theta),\mathbf{y})}{(\mathbf{x},\mathbf{y})\sim \mathcal{D}} \;,
\end{align}
where $\delta^{*}={\argmax}_{\delta, \left\|\delta\right\|_\infty\le \varepsilon}J(f(g(\mathbf{x}+\delta,\phi),\theta),\mathbf{y})$.
\subsection{Adversarial fine-tuning of joint ASR and Denoiser model with ASR model frozen} 
It is known that adversarial fine-tuning may over-fit the ASR model to work well on adversarial examples degrading benign performance. To avoid this, we tried another variant of the joint network by freezing the ASR model and just updating the denoiser parameters. We call this model 
\textit{ADV-FINETUNE-JOINT-ASRfrozen}. We expect that adversarial training on the denoiser will be more efficient than doing it on the  full ASR+denoiser network as the denoiser network is much smaller than ASR and hence will require less epochs to train.
\begin{table}
\centering
\caption{\label{tab:wer_results2}
Word error rate (\%) for K2 ASR system under PGD-500 attacks.  [Ground-truth WER (GT) and target WER (TGT). \textit{RS$\sigma$} stands for randomized smoothing with a $\sigma$ parameter. Arrow  $\downarrow$ indicates lower is better and  $\uparrow$ indicates higher is better.}
\resizebox{0.47\textwidth}{!}{
\begingroup
\setlength{\tabcolsep}{3pt}
\begin{tabular}{@{}lccccc@{}}
\toprule
\textbf{System} & \textbf{Benign} & \multicolumn{4}{c}{\textbf{PGD-500 Attack}} \\
\cmidrule(rl){2-2}
\cmidrule(l){3-6}
$L_{\infty}$ (max-norm) &  &  \multicolumn{2}{c}{0.01} &   \multicolumn{2}{c}{0.1}  \\
\cmidrule(lr){3-4} \cmidrule(l){5-6}
& GT $\downarrow$  & GT $\downarrow$ & TGT $\uparrow$ &   GT $\downarrow$ & TGT $\uparrow$ \\
\midrule
Baseline (reduced LibriSpeech test-clean) &  \bf 4.53 & 97.81 & 40.71   &  100.59 &	16.47   \\
Baseline + \textit{RS0.001} &  4.78 &  101.41 & 16.37  & 101.22 & 17.19\\
Baseline + \textit{RS0.001} + \textit{Denoiser} & 4.63   & 94.05 & 55.40  & 92.44 & 71.96  \\
\midrule
\textit{ADV-FINETUNE-ASR} + \textit{RS0.001}  & 4.97 & 68.84 & 91.07 &  101.95 &	10.03 \\
\midrule
 \multirowcell{2}[0pt][l]{\textit{ADV-FINETUNE-JOINT} \\
 \hspace{2em} + \textit{RS0.001}} & \multirowcell{2}[0pt][c]{5.17} & \multirowcell{2}[0pt][c]{62.36} & \multirowcell{2}[0pt][c]{97.94}  &  \multirowcell{2}[0pt][c]{96.25} & \multirowcell{2}[0pt][c]{48.92} \\
 &&&&&\\
 \midrule
 \multirowcell{2}[0pt][l]{\textit{ ADV-FINETUNE-JOINT-ASRfrozen} \\
 \hspace{2em} + \textit{RS0.001}} & \multirowcell{2}[0pt][c]{5.70} & \multirowcell{2}[0pt][c]{\textbf{25.40}} & \multirowcell{2}[0pt][c]{\textbf{100.05}}  &  \multirowcell{2}[0pt][c]{\textbf{82.84}} & \multirowcell{2}[0pt][c]{\textbf{93.23}} \\
 &&&&&\\
\bottomrule
\end{tabular}
\endgroup
}
\end{table}

\section{Experimental Setup}
\label{sec:exp}
\textbf{Dataset, Baseline, and Adversarial Attacks}
\label{ssec:dataAdv}
Our experimental setup was based on LibriSpeech dataset~\cite{panayotov2015librispeech}. 
We used a Hybrid DNN-HMM ASR model implemented on the K2 framework\footnote{\url{https://k2-fsa.github.io/k2/index.html}} using Pytorch~\cite{paszke2019pytorch}.
A Conformer~\cite{gulati2020conformer} network was used to compute frame-level posteriors, which were used as input to the K2 WFST decoder. The Conformer consisted of 12-layers with dimension=256, heads=4, and feed-forward dimension=2048; and used 80 log-Mel-filterbank features. The whole pipeline is end-to-end differentiable to be able to compute adversarial examples in time domain.
This model was trained 
on the full 960 hours LibriSpeech corpus for 20 epochs, and the parameters from the last 5 epochs were averaged to get the final model. 
We denote this model as the undefended \textit{baseline}. 
To evaluate the robustness of the model, we applied targeted FGSM and PGD attacks with different strengths, i.e., max. $L_\infty$ norm levels: \{0.0001, 0.001, 0.01, 0.1, 0.2\}. For PGD, we evaluated attacks with 7 and 500 iterations (PGD-7 and PGD-500). The target phrases used to craft the attacks were taken from the LibriSpeech training set. For each utterance, we chose a target phrase with length similar to the benign transcript. 
We evaluated the attacks on the first 100 examples from LibriSpeech test-clean. We chose to work on this reduced set because of three reasons. First, the computational cost for PGD iterations is too high for the full set, so in the literature, it is common  to experiment on a limited number of utterances~\cite{carlini-18}. Second, for the baseline, the performance of the reduced set did not statistically change w.r.t. the full set 
(see Table~\ref{tab:wer_results}). Lastly, this setup is the same as in previous work~\cite{zelasko2021adversarial} and in DARPA-GARD evaluations. Another important thing to point-out is that we always evaluated using white-box adaptive attacks. That means that the adversary knows the defenses and their parameters and it is able to back-propagate through them to create the adversarial examples. Many works in the adversarial literature do not consider adaptive attacks based their defense in obfuscating the gradients of the system as evidenced in~\cite{athalye2018obfuscated}.
\vspace{1mm} \\
\textbf{Attacks Dataset:} We create a dataset by generating offline PGD adversarial samples against LibriSpeech train sets using $L_2=\{0.2,0.5,1.5,1.9\}$ and $L_\infty=\{0.001,0.01,0.1\}$ threat models with number of iterations $\{10,20,50,100,200\}$ sampled with more bias towards high norm and high iteration attacks. Generating offline adversarial examples has the advantage that we can use computationally expensive PGD attacks with large number of iterations without letting a model run for months on online attacks. The generated attacks were used to train the denoiser as described in Section~\ref{ssec:denoi}. 
Then, the pre-trained denoiser could be finetuned in tandem with ASR model to increase its robustness.
\vspace{1mm} \\
\textbf{Denoiser:}
\label{ssec:denoi}
After experimenting with a few denoiser architectures, we choose TasNet~\cite{luo2019conv}, a time-domain model for source separation and speech enhancement.
It is an all-convolutional 1-D Convolutional Neural Network (CNN), which consists of 
\emph{encoder}, \emph{separator}, and \emph{decoder}.
The \emph{encoder} and \emph{decoder} are single convolutional layers, while the \emph{separator} stage consists of multiple CNNs called \emph{stacks}.
The \emph{stacks} output are combined to produce a mask which is applied to the encodings and passed to the \emph{decoder} stage. We used 128-dim encodings obtained with kernel-size=16 and stride=8. The separator used one stack with 16 layers with kernel dilations increasing with a factor of 2~\cite{luo2019conv}.
We trained the denoiser on our dataset of offline attacks using the corresponding benign example as clean target.
\vspace{1mm} \\
\textbf{\textit{ADV-FINETUNE-ASR:}}
We fine-tuned the baseline ASR model on on-the-fly PGD-7 attacks.
The $L_\infty$ for this attacks were randomly sampled from a log uniform distribution [0.0001,0.02]. The learning rate was 10x lower than the one used in the training phase.
\vspace{1mm} \\
\textbf{\textit{ADV-FINETUNE-JOINT:}}
Instead of adversarially fine-tuning just the ASR model, we jointly fine-tuned the tandem denoiser+ASR. 
We evaluated another variant where attacks were generated on the tandem denoiser+ASR but we only updated the denoiser weights while the ASR model is frozen. 
We call this defense as \textbf{\textit{ADV-FINETUNE-JOINT-ASRfrozen}}.
\section{Results}
\label{sec:results}

We evaluate the robustness of baseline ASR and all proposed defenses against FGSM and PGD-7 attacks (Table ~\ref{tab:wer_results}) and PGD-500 attacks (Table ~\ref{tab:wer_results2}).
For FGSM and PGD-7, the WER w.r.t. the target phrase was always greater than 90\%.  In other words, 
the adversary is not able to make the ASR to recognize the malicious target phrase. 
Hence we omitted TGT WER in Table~\ref{tab:wer_results} and included only WER w.r.t ground truth phrase (GT WER). We analyze the undefended baseline 
for full and reduced Librispeech test-clean set. We observe that the results for both sets are similar. Therefore, henceforward, all models were evaluated on the reduced set to alleviate the large cost of generating attacks
(and other reasons mentioned in Section~\ref{sec:exp}). Next, we evaluated the different defenses.
We observe that the denoiser defense trained on offline attacks performed better than randomized smoothing for most $L_\infty$ values and on par with the adversarially trained ASR. Both defenses jointly adv. fine-tuning denoiser and ASR performed significantly better than the offline denoise and the adv. fine-tuned ASR. 
\textit{ADV-FINETUNE-JOINT} yielded the largest robustness against FGSM attacks with mean absolute decrease of 19.3\% GT WER w.r.t. the baseline.
\textit{ADV-FINETUNE-JOINT-ASRfrozen} was the best for PGD-7 with a mean absolute decrease in GT WER of 22.3\% w.r.t. the baseline. We can observe that for both attacks, the best defense kept the system robust up to $L_\infty\le 0.01$.

When increasing the number of PGD iterations to 500 (Table~\ref{tab:wer_results2}), the WER w.r.t. the target phrase (TGT) decreases, meaning that the attacker starts being sucessful in making the system to recognitize a particular malicious phrase. However, the obtain an usable TGT WER of 16\%, it needs to increase $L_\infty$ to 0.1, which is a very perceptible attack. Here, the goal of the defense is increasing TGT WER while reducing GT WER. Again, the best system by far was \textit{ADV-FINETUNE-JOINT-ASRfrozen}.
The mean absolute decrease GT WER was 45.08\% GT WER and increase in TGT WER was 68.05\%. Although there was a slight increase (1.17\%) in the benign WER  with reference to the baseline model, the gain adversarial robustness overshadowed it.
The proposed method significantly outperformed the baseline defenses in the literature, i.e., randomized smothing and ASR adversarial training. Unfortunately, for large $L_\infty=0.1$, the defenses could not reduce GT WER much. However, note that these are very perceptible attacks and even using Gaussian noise (non adversarial) of that level would significantly damage the ASR system.
\begin{figure}
\centering
\includegraphics[trim=0.6cm 0cm 0cm 0cm,clip=true,width=1\linewidth]{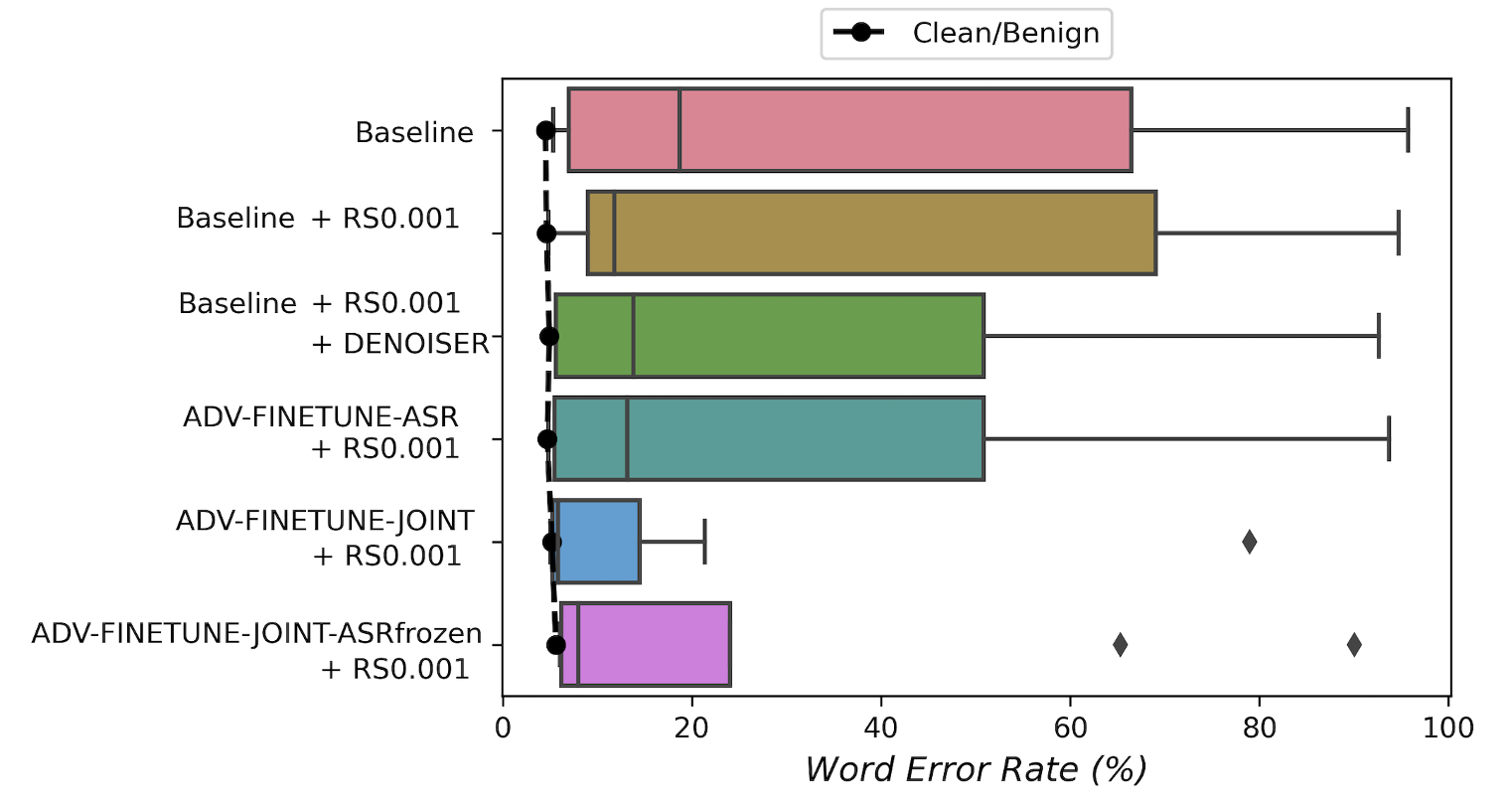}
\caption{Summary of systems for all attack settings using boxplot. We exclude FGSM-$L_{\infty}=0.2$ and PGD-$L_{\infty}=0.2$) because they are perceptible attacks with SNR . The dotted line indicates the clean/benign performance of the system}
\label{fig:boxplot}
\end{figure}

\section{Conclusion}
\label{sec:concl}
We evaluated the robustness of K2 Hybrid ASR model alongwith four defenses--pre-processing time-domain denoiser defense, adversarial fine-tuning of ASR model and two variants of joint adversarial training of pre-processing denoiser and ASR model. We evaluated these defenses against strong adaptive white-box attacks i.e. when the adversary is aware of parameters of defense model along with those of ASR. To understand the big picture of the defenses, we convert the Table ~\ref{tab:wer_results} to boxplot shown in Figure \ref{fig:boxplot}. Please note that we exclude FGSM-$L_{\infty}=0.2$ and PGD-$L_{\infty}=0.2$) as they are perceptible. The dotted blue line indicates the clean/benign performance of the individual systems. The best defense should have WER distribution concentrated around the blue line. The results show that \textit{ADV-FINETUNE-JOINT} is the best defense, followed closely by \textit{ADV-FINETUNE-JOINT-ASRfrozen}. On the other hand, for PGD-500 attacks, \textit{ADV-FINETUNE-JOINT-ASRfrozen} performs the best, consistently yielding low GT WER and high TGT WER. This is at the cost of slight degradation in benign WER ($<$1.2\%), however this is expected for adversarial defenses. In the future, we would like to bridge this gap furthur.

\clearpage
\bibliographystyle{IEEEtran}
\bibliography{mybib}
\end{document}